\documentclass{PoS}

\title{HAWC Upgrade with a Sparse Outrigger Array}

\ShortTitle{HAWC Upgrade with a Sparse Outrigger Array}

\author{\speaker{Andres Sandoval}$^a$ , for the HAWC Collaboration$^b$ \\
        \llap{$^a$}Instituto de Fisica, Universidad Nacional Autonoma de Mexico, Mexico City, Mexico \\
         \llap{$^b$}For a complete author list, see \href{http://www.hawc-observatory.org/collaboration/icrc2015.php}{www.hawc-observatory.org/collaboration/icrc2015.php}\\
        Email: \email{asandoval@fisica.unam.mx}}

\abstract{The High Altitude Water Cherenkov (HAWC) high-energy gamma-ray observatory has recently been completed on the slopes of the Sierra Negra volcano in central Mexico. HAWC consists of 300 Water Cherenkov Detectors, each containing 180 m$^3$ of ultra-purified water, that cover a total surface area of 20,000 m$^2$. It detects and reconstructs cosmic- and gamma-ray showers in the energy range of 100 GeV to 100 TeV. The HAWC trigger for the highest energy gammas reaches  an effective area of 10$^5$ m$^2$ but many of them are poorly reconstructed because the shower core falls outside the array. An upgrade that increases the present fraction of well reconstructed showers above 10 TeV by a factor of 3-4 can be done with a sparse outrigger array of small water Cherenkov detectors that pinpoint the core position and by that improve the angular resolution of the reconstructed showers. Such an outrigger array would be of the order of 200 small water Cherenkov detectors of 2.5 m$^3$  placed over an area four times larger than HAWC. Detailed simulations are being performed to optimize the layout.}

\FullConference{The 34th International Cosmic Ray Conference,\\
		30 July- 6 August, 2015\\
		The Hague, The Netherlands}
\usepackage{lineno}
\begin{document}
\section{Introduction}

Our knowledge of the gamma-ray sky has greatly been expanded by the FERMI satellite data, but at the highest energies above 1 TeV only 161 sources are known up to date \cite{TVCAT:2015}. The first one corresponds to the observation by the Whipple observatory of the Crab supernova remnant in 1989 \cite{Weekes:1989}. Most of the other sources have been discovered recently by the Imaging Atmospheric Cherenkov (IAC) telescopes VERITAS \cite{VERITAS},  HESS \cite{HESS} and MAGIC \cite{MAGIC} (see also \cite{ICRC-2015} in these proceedings ).

The gamma-ray sky at the highest energies above 10 TeV is even less well studied but is of great interest since sources  that emit gammas at these energies are most probably associated to the PeVatrons that accelerate the high energy cosmic rays and produce the astrophysical high energy neutrinos detected recently by IceCube \cite{IceCube}. Diffuse emission or extended sources at the highest energies could be indirect signatures of dark matter \cite{DM} or of more exotic phenomena \cite{Exotic}. It is therefore desirable to have a detector able to do a systematic survey of a large region of the sky with enough sensitivity above 10 TeV.

From 2000 to 2008 the Milagro observatory  \cite{MILAGRO} operated in the Jemez mountains in New Mexico demonstrating that large area water Cherenkov detectors could be used to detect the shower particles as they hit the ground, reconstructing them and identifying which ones  originated from gamma rays. Milagro discovered several galactic sources both point-like and extended.

HAWC is a large aperture continuously operating observatory that  scans 2/3 of the celestial sphere each day detecting high-energy showers from gamma and cosmic rays, reconstructing their direction and energy. Over the 10 years of planned operation, HAWC will will make a  systematic survey of galactic and extragalactic gamma-ray sources in this region, complementing the observations of HESS, VERITAS and MAGIC.

\section{The HAWC Observatory}
 
\begin{figure}[h]
\centering
{\includegraphics[height=7cm]{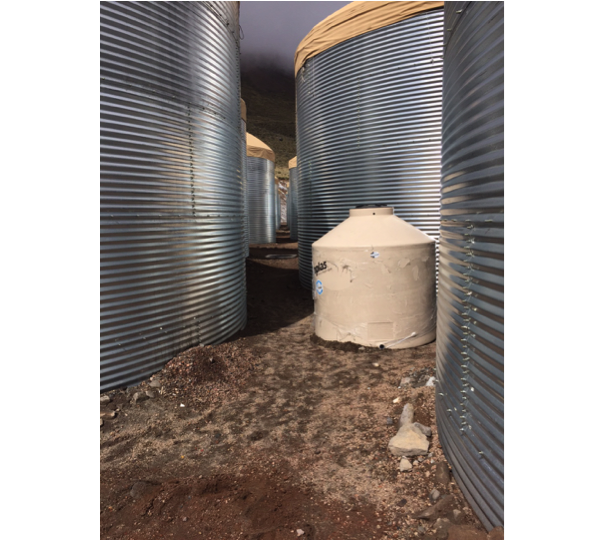}}
\caption{Photograph of some of the 300 WCDs of the HAWC observatory having each 180,000 liters of water. A prototype of an outrigger tank of 2,500 l is also seen.}
\end{figure}

The HAWC observatory  \cite{HAWC:2015}  is a second-generation water Cherenkov detector based on the detection technique developed by  Milagro. It has been built on the Sierra Negra volcano in central Mexico at 4,100m above sea level. HAWC consists of an array of 300 closely packed cylindrical water Cherenkov detectors (WCD) of 7.3m diameter and 4.5m high (Figure 1). They are filled with high purity water and contain each 4 photomultipliers (PMT) anchored at the bottom. The PMTs  detect the Cherenkov light produced in the water by the arriving particles from the air showers. The array covers an area of 20,000 m$^2$ in a tightly packed geometry. 

All the signals in the PMTs  are recorded using the Time over Threshold (ToT) technique to determine both the arrival time and the amplitude of the signals without dead time. Taking data continuously without a trigger, the digitized signals constitute a data volume of close to 500 MB/s. This data stream is inspected by a farm of online computers which make a software event selection to reduce the event recording rate on disk to 20 MB/s. Presently this corresponds to a 23 KHz rate of showers having  more than 28 PMTs  with  signals in coincidence in a 150 ns time window.

For each event the shower front is reconstructed using the arrival time of the signals of the PMTs of the different tanks, that are measured with a time resolution better than 1ns. The shower front is first fitted with a plane followed by a fit with the curvature corrections and shower time spread given by the actual propagation of particles in the atmosphere. The position of the shower core is determined by fitting a Nishimura-Kamata-Greisen (NKG) distribution to the amplitude of the detected signals \cite{NKG}. This can reliably be done if the core falls inside the array (20,000 m$^2$ effective area) but becomes more ambiguous if the core falls outside of the array.

\begin{figure}[h]
\centerline{\includegraphics[height=10cm]{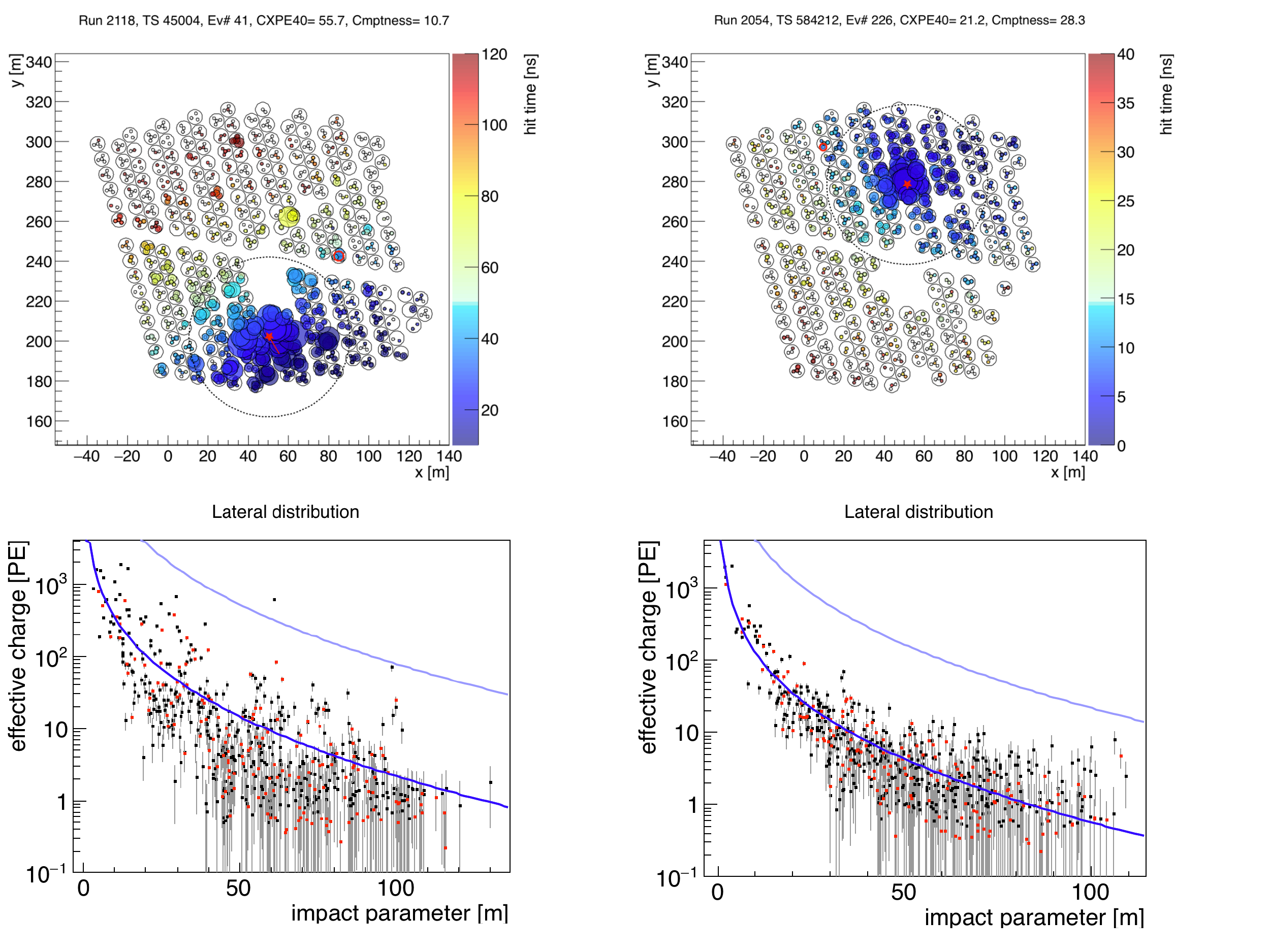}}
\caption{HAWC events of a hadronic shower (left) and an electromagnetic one (right) showing the PMT signals on the array  (top row) and as function of the distance from the shower core with the NKG fit (bottom row).}
\end{figure}

The gamma/hadron discrimination is performed by the topology and pulse height distribution in the event as function of the distance to the shower core (Figure 2). This method performs better as the energy of the gamma increases and at the highest energies the events are almost background free. 

Nevertheless there is a large fraction of showers in the trigger where the core falls outside of the HAWC array, that leave enough information in the WCDs to do a gamma/hadron discrimination, but for which there is an ambiguity in the distance of the core position, the direction of the shower and the shower size. To recover a large fraction of these showers specially at the highest energies, where we are already limited by the statistics, is a desirable upgrade program for the HAWC observatory.

\section{Outrigger arrays}
  
The Milagro instrument consisted initially of a 4000 m$^2$ pond. After 3 years of operation a sparse outrigger array was added. This increased dramatically the sensitivity of the observatory by being able to determine the shower core position over an area much bigger than the pond and thereby correctly reconstructing partially detected showers.

The aim of the HAWC outrigger array is to determine the position of the shower core for showers falling outside the HAWC WCD array and that still leave enough information in HAWC to reconstruct the shower front and discriminate between gamma and cosmic-ray initiated showers. This naturally limits the improvement to large energy showers above ~1 TeV. The dimensions of a high-energy shower footprint at the HAWC altitude sets a limit on the maximum outrigger radius at which it is still efficient to deploy the small tanks. The required resolution of the shower core position will dictate the density of detectors. The fact that the outriggers will be measuring particles close to the shower core for large showers means that there will be large signals and therefore smaller WCDs can be used.

In order to optimize the outrigger geometry, detailed Monte Carlo simulations are being performed. A general outrigger geometry with 300 tanks arranged in a sunflower spiral manner is used with CORSIKA 
\cite{CORSIKA} and GEANT4 \cite{GEANT4} simulation packages. To study different sizes of the outrigger array, any of the 300 tanks can be switched off at the event reconstruction level.

\begin{figure}[h]
\centerline{\includegraphics[height=10cm]{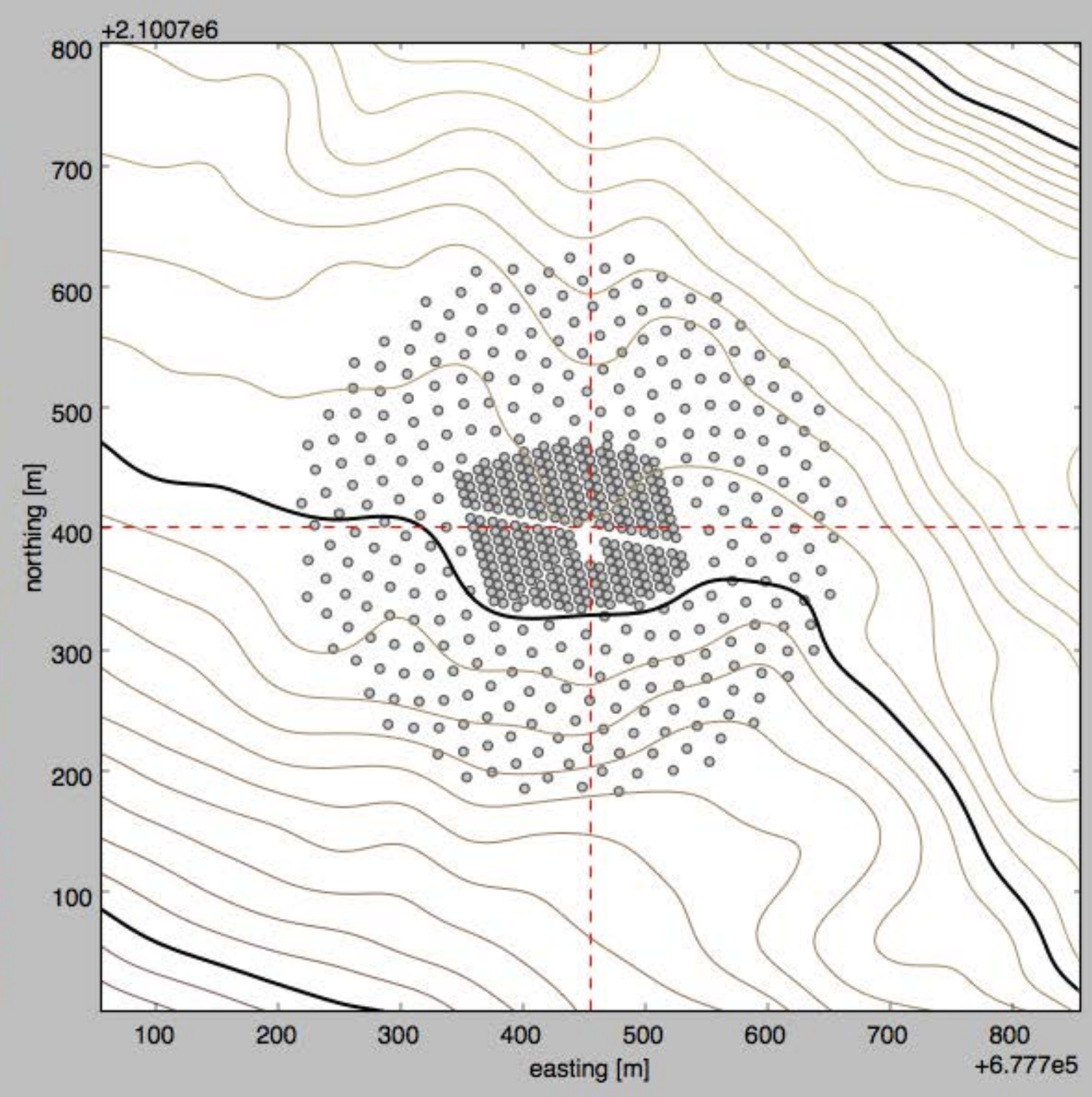}}
\caption{Contour lines of the terrain at the HAWC site with the HAWC array and the layout of a 300 tank outrigger array in the geometry of a sunflower spiral.}
\end{figure}

The outriggers can be chosen from different readily available commercial water tanks in Mexico having water volumes of 2.5 to 3.5 m$^3$. Different types of the inner surfaces are being simulated, from black absorbing to white diffuse reflecting. We are also evaluating 3$"$ and 8$"$ PMTs both anchored at the bottom of the tank and looking up like in HAWC and at the top and looking down as in Milagro outriggers and Auger.

We expect, based on our prior experience with  Milagro, that an outrigger array will boost the effective area by a factor of 3 to 4, with the largest gain above 10 TeV.



Besides the design of the small WDCs and the layout of the outrigger array geometry it is necessary to define the Front End Electronics (FEE), the HV power supplies and the data acquisition system. It is desirable to place all of them locally on the individual outrigger WDCs. The HV will be of the  DC-DC converter type. The FEE will consist of a  shaper and a leading edge discriminator to trigger the FADC converter digitizing the signal shape that is readout by a small processor and sent via optical fibre to the main HAWC data acquisition system.  We plan to use the outriggers to test new local digital electronics systems, so the tanks can act both as an upgrade to HAWC and as a prototype for a possible HAWC-South array \cite{M.Duvernois}.

\section{Conclusions}

An upgrade of the HAWC high-energy gamma-ray observatory with a sparse array of small outrigger tanks is being investigated. For > 10 TeV showers  HAWC has a trigger effective area that is much larger than its physical size, but for showers where the core falls outside the array there are ambiguities in the reconstruction between the core position, the shower angle and the shower size or energy. An outrigger array can determine the core position for showers falling outside the main array elevating the ambiguities and making these showers well reconstructable. A gain of 3-4 in sensitivity for gammas above 10 TeV can be obtained over what is presently achieved.

\section*{Acknowledgments}
\footnotesize{
We acknowledge the support from: the US National Science Foundation (NSF);
the US Department of Energy Office of High-Energy Physics;
the Laboratory Directed Research and Development (LDRD) program of
Los Alamos National Laboratory; Consejo Nacional de Ciencia y Tecnolog\'{\i}a (CONACyT),
Mexico (grants 260378, 232656, 55155, 105666, 122331, 132197, 167281, 167733);
Red de F\'{\i}sica de Altas Energ\'{\i}as, Mexico;
DGAPA-UNAM (grants IG100414-3, IN108713,  IN121309, IN115409, IN111315);
VIEP-BUAP (grant 161-EXC-2011);
the University of Wisconsin Alumni Research Foundation;
the Institute of Geophysics, Planetary Physics, and Signatures at Los Alamos National Laboratory;
the Luc Binette Foundation UNAM Postdoctoral Fellowship program.
}


\end{document}